\documentstyle[12pt,fleqn]{article}
\def\mpl{{{m_{Pl}}}}
\def\sig{{\sigma}}
\def\lam{{\lambda}}
\def\ep{{\epsilon}}

\def\eps{{\epsilon}}
\def\alp{{\alpha}}

\def\del{{\delta}}
\def\gam{{\gamma}}
\def\bet{{\beta}}

\def\znew{{\sigma}}
\def\be{\begin{equation}}
\def\ee{\end{equation}}
\def\ba{\begin{eqnarray}}
\def\ea{\end{eqnarray}}
\def\la{\mathrel{\mathpalette\fun <}}
\def\ga{\mathrel{\mathpalette\fun >}}
\def\fun#1#2{\lower3.6pt\vbox{\baselineskip0pt\lineskip.9pt
        \ialign{$\mathsurround=0pt#1\hfill##\hfil$\crcr#2\crcr\sim\crcr}}}

\def\re#1{{[\ref{#1}]}}
\def\eqr#1{{Eq.\ (\ref{#1})}}
\begin{document}
\begin{titlepage}
\null\vspace{-62pt}
\begin{flushright}
{\footnotesize
FERMILAB--Pub--94/046-A\\
CfPA 94-th-14\\
astro-ph/9403001\\
February 1994 \\
Submitted to {\em Phys. Rev. D}}
\end{flushright}
\renewcommand{\thefootnote}{\fnsymbol{footnote}}
\vspace{0.5in}

\begin{center}
{\Large \bf Relating spectral indices to tensor\\ and scalar amplitudes in
inflation}\\
\vspace{1.0cm}
Edward W.\ Kolb\footnote{Electronic mail: {\tt rocky@fnas01.fnal.gov}}\\
{\em NASA/Fermilab Astrophysics Center\\
Fermi National Accelerator Laboratory, Batavia, IL~~60510, and\\
Department of Astronomy and Astrophysics, Enrico Fermi Institute\\
The University of Chicago, Chicago, IL~~ 60637}\\
\vspace{0.4cm}
Sharon L.\ Vadas\footnote{Electronic mail: {\tt
vasha@physics.berkeley.edu}}\\
{\em Center for Particle Astrophysics\\
University of California, Berkeley, CA~~94720}\\
\end{center}

\baselineskip=24pt

\vspace{1.5cm}

\begin{quote}
\hspace*{2em}  Within an expansion in slow-roll inflation parameters,
 we derive the complete second-order expressions relating the ratio of
tensor to
scalar density perturbations and the spectral index of the scalar
spectrum.  We find that ``corrections'' to previously derived formulae
can dominate if the tensor to scalar ratio is small.  For instance, if
$V V''/(V')^2\neq 1$ or if $m_{Pl}^2/(4\pi) ~|V'''/V'|\ga 1$, where
$V(\phi)$ is the inflaton potential and $m_{Pl}$ is the Planck mass,
then the previously used simple relations between the indices and the
tensor to scalar ratio fails. This failure occurs in particular for
natural inflation, Coleman--Weinberg inflation, and ``chaotic''
inflation.
\vspace*{12pt}

PACS number(s): 98.80.Cq, 04.30.$+$x, 98.70.Vc

\renewcommand{\thefootnote}{\arabic{footnote}}
\addtocounter{footnote}{-2}
\end{quote}
\end{titlepage}
\addtocounter{page}{1}
\newpage
\baselineskip=24pt
\baselineskip=14pt

\vspace{36pt}
\centerline{\bf I. INTRODUCTION}
\vspace{24pt}

In slow-roll inflation the energy density of the Universe is dominated
by the potential energy density of some scalar field $\phi$, known as
the {\em inflaton} field. During slow-roll inflation, scalar density
perturbations and gravitational mode perturbations are produced as the
inflaton field evolves.  The amplitude of the scalar density
perturbation as it crosses the Hubble radius after inflation is
defined as
\be
\left(\frac{\delta \rho}{\rho}\right)_\lambda^{\rm HOR} \equiv
\frac{m}{\sqrt{2}} A_S(\lambda),
\ee
where the constant $m$ equals $2/5$ (or $4$) if the perturbation
re-enters during the matter (or radiation) dominated era.  In addition
to the scalar density perturbations, slow-roll inflation produces
metric fluctuations, $h$, and the amplitude of the dimensionless
strain on scale $\lambda$ when it crosses the Hubble radius after
inflation is defined by
\be
\left|k^{3/2}h\right|_\lambda^{\rm HOR} \equiv A_G(\lambda).
\ee
Of particular interest is the ration of the tensor to scalar perturbations,
defined as
\be
R(\lambda) \equiv \frac{A_G^2(\lambda)}{m^2A_S^2(\lambda)/2}
\ee

Both the scalar density perturbations and the tensor modes contribute
to temperature fluctuations in the cosmic background radiation (CBR).
On large angular scales ($\theta\gg1^\circ$, corresponding to the
horizon at the last scattering surface) CBR fluctuations are
proportional to the sum of the squares of the two
modes:\footnote{There are three interrelated scales used to
characterize sizes: $\lambda$, $\theta$, and $\phi$.  The length scale
$\lambda$ is related to the angular scale $\theta$ by $\lambda =
\theta (34.4^{\prime\prime}\Omega_0h)^{-1}$Mpc; i.e., $\theta$ is the
angle subtended on the sky today by comoving scale $\lambda$ at the
surface of last scattering of the CBR. This comoving scale $\lambda$
crossed the Hubble radius during inflation when the value of the
scalar field was $\phi$.}
\be
\left[\Delta T(\theta)/T\right]^2 \propto S(\theta) + T(\theta),
\ee
where with the normalization above,\be
\label{eq:ST}
S(\theta)=\frac{m^2}{2}A_S^2(\phi); \qquad T(\theta)=A_G^2(\phi).
\ee
Here we will be interested in scales that crossed the Hubble radius
during the matter-dominated epoch when $m=2/5$.

It is convenient to parameterize the scalar and tensor spectrum by their
spectral indices:
\ba
\label{eq:defnS}
1-n_S  & \equiv &   d \ln \left[m^2 A_S^2 (\lambda)/2\right]/ d \ln \lambda
= d \ln [S (\lambda)]/ d \ln \lambda \nonumber  \\
\label{eq:defnT}
n_T & \equiv & -d \ln [A_G^2 (\lambda)]/ d \ln \lambda
= -d \ln [T (\lambda)]/ d \ln \lambda,
\ea
where $\lambda$ is the physical wavelength that a given scale would
have today if it evolved linearly, and is given by
$\lambda=[a(t_0)/a(t)]~H^{-1}(t)$, where $t_0$ is the time today, and
$a(t)$ and $H(t)=\dot{a}(t)/a(t)$ are the scale factor and Hubble
expansion rate, respectively, when the scale left the Hubble radius at
time $t$.

With the prospect of measurements of anisotropy in the cosmic
microwave background at different angular scales, it may soon be
possible to determine the relative contributions of scalar and tensor
components to CBR fluctuations \re{BUNCH}, and thus provide
information about the scalar potential driving inflation \re{CKLL}.
Several attempts have already been made to develop a method to isolate
the scalar and tensor components of the signal \re{CBDES}.  This work
assumed a relationship between the ratio of the tensor and scalar
contribution to the temperature fluctuations and the spectral indices
of the form
\ba
\label{eq:tsntold}
T/S  & \simeq & -7n_T \simeq 7(1-n_S+\zeta) \\
\label{eq:tsold}
     & \simeq & 7(1-n_S),
\ea
where $\zeta\equiv [2V'/(3H^2)]'$, $V$ is the inflaton potential,
prime denotes $d/d\phi$, and $H$ is the expansion rate at the time
when the scale $\lambda$ crossed out of the Hubble radius during
inflation. It is claimed in Ref.\ \re{CBDES} that $\zeta$ is small in
generic models of inflation, and that confirmation of $n_T \approx
n_S-1$ would be support for inflation and provide detailed information
about the first instants of the Universe.  In this paper, we discuss
the relationship between $R$, $n_T$ and $1-n_S$ within an expansion of
the scalar and tensor amplitudes in terms of ``slow-roll''
parameters.\footnote{We derive these relations for the density
fluctuation amplitudes rather than the actual measured temperature
fluctuations.  This will account for the difference between $6.25$ and
$7$ in the expressions for $1-n_S$ and $n_T$.} We confirm Eq.\
({\ref{eq:tsntold}}) as the lowest-order result in this expansion if
$|dR/d\ln\lam|\gg|d^2R/d\ln\lam^2|$, and we derive the next-order
terms in the expansion as well.  We also discuss when the
approximation in Eq.\ ({\ref{eq:tsold}}) is accurate, i.e., what is
the relative magnitude of the $\zeta$ contribution compared to
$1-n_S$.

We will now discuss the equations relating the tensor and scalar
amplitudes to the spectral indices.  In the Hamilton--Jacobi treatment
of the field equations for inflation, the scalar field $\phi$ is used
as a time variable, and the field equations are
\re{HJ}
\be
\label{eq:eom}
[H'(\phi)]^2 - \frac{3}{2} \kappa^2 H^2(\phi) =  - \frac{1}{2} \kappa^4
V(\phi); \qquad
\kappa^2 \dot{\phi}  =  -2 H'(\phi) ,
\ee
where dot denotes a time derivative, and $\kappa^2 =8 \pi/m^2_{Pl}$
with $m_{Pl}$ the Planck mass.  The scalar and tensor perturbations
are calculated in an expansion in slow-roll parameters.  These
parameters involve combinations of derivatives of $H(\phi)$.  In de
Sitter space $H(\phi)$ is a constant ($\dot{\phi}=0$).  However, in
slow-roll inflation $H(\phi)$ varies with time, albeit slowly.  The
familiar first-order result for $A_G$ and $A_S$ is
\be
\label{eq:ASAG}
A_S(\phi) = -  \frac{\sqrt{2} \kappa^2}{8\pi^{3/2}}\,
\frac{H^2(\phi)}{H'(\phi)};
	\qquad
A_G(\phi) = \frac{\kappa}{4\pi^{3/2}} \, H(\phi)  .
\ee
However there are corrections to this result that depend upon the
slow-roll expansion parameters $\epsilon$ and $\eta$, which are
defined as \re{LiddLyth}
\ba
\label{eq:ep}
\epsilon(\phi) & = & \frac{2}{\kappa^2} \left[ \frac{H'(\phi)}{H(\phi)}
\right]^2, \\
\label{eq:eta}
\eta(\phi) & = & \frac{2}{\kappa^2} \, \frac{H''(\phi)}{H(\phi)} =
\ep+\frac{H(\phi)}{2H'(\phi)}\ep'.
\ea
These parameters depend upon $H'^2(\phi)$ and $H''(\phi)$; they are
second order in derivatives.  In the slow-roll approximation $\ep$ and
$\eta$ are less than one.  A third parameter $\xi$, defined as
\be
\label{eq:xi}
\xi(\phi) = \frac{2}{\kappa^2} \, \frac{H'''(\phi)}{H'(\phi)} =
\eta + \frac{H(\phi)}{H'(\phi)}\eta',
\ee
depends upon $H'''(\phi)/H'(\phi)$, and will appear in the expression
for the scalar spectral index.  The parameter $\xi$ can be much larger
than one, even if $\epsilon$ and $\eta$ are small.\footnote{Even if
$\eta$ is smaller than one, the derivative of $\eta$ can be large,
resulting in $|\xi|>1$.  This turns out to be the case for a wide
range of parameters in Coleman-Weinberg inflation.}

The expressions for $A_G$ and $A_S$ in Eq.\ ({\ref{eq:ASAG}}) are
correct to first order in $\{\epsilon,\ \eta\}$, and we will refer to
this as the ``first-order'' results. To second order, Stewart and Lyth
showed \re{SL}:
\ba
\label{eq:AS}
A_S(\phi)& = & -  \frac{\sqrt{2} \kappa^2}{8\pi^{3/2}} \,
\frac{H^2(\phi)}{H'(\phi)} \, \left[ 1 - (2C+1)\epsilon +
C\eta \right] \\
\label{eq:AG}
A_G(\phi) & = & \frac{\kappa}{4\pi^{3/2}}\, H(\phi)\, \left[ 1-(C+1)\epsilon
\right] ,
\ea
where $C\equiv -2+\ln 2 +\gamma\simeq -0.73$ and $\gamma=0.577$ is
Euler's constant.  In deriving this expression, terms of order
$\ep^2$, $\ep\eta$ and $\eta^2$, etc., have been neglected in the
square bracket.  It is important to realize that Eq.\ ({\ref{eq:AS}})
is a double expansion in $\epsilon$ and $\eta$, and that they might
not be of the same order of magnitude.  It makes sense to include all
terms in the square brackets of Eq.\ (\ref{eq:AS}) as long as
$|\eta|\sim \ep$.  However, because $\eta$ depends on the derivative
of $\ep$, this is usually not the case.  It is possible that the
second-order terms in one expansion variable might be larger than the
first-order terms in the other variable.  We will see later that this
occurs for many inflationary models of interest.

Let us now find the expressions for $R$, $1-n_S$, and $n_T$ to second
order in the slow-roll parameters.  We can find $R$ directly from Eqs.\
({\ref{eq:AS}}) and ({\ref{eq:AG}}).  Using the relation
[\ref{CKLL},\ref{CKLL3}]
\be
\label{eq:dlamdphi}
\frac{d\ln\lambda}{d\phi} = \frac{\kappa^2}{2} \frac{H}{H'} \left[ 1-\epsilon
\right],
\ee
we can also find the spectral indices.  The complete second-order
expressions then are
\ba
\label{eq:TSgen}
R & = & \frac{25}{2}\, \frac{A_G^2}{A_S^2}
\simeq \frac{25}{4}\, 2\ep\left[1-2C(\eta-\ep)\right]\\
\label{eq:nT}
n_T & = & -2\eps\left[ 1+(2C+3)\ep-2(C+1)\eta \right ] \\
\label{eq:nsgen}
1-n_S & = & 2\ep\left[2-\frac{\eta}{\ep}+4(C+1)\ep-(5C+3)\eta + C\xi
(1+2(C+1)\ep-C\eta)
\right].
\ea
As mentioned above, $\xi$ can be of order or greater than one, and is
therefore not an expansion variable as are $\ep$ and
$\eta$.\footnote{Although $|\xi|$ can be much larger than one, it must
be less than $1/\ep$: $|\xi|\ll 1/\ep$.  This is because $|\eta
(d{\ddot{\phi}}/dt)/(H\ddot{\phi})|= |\xi\ep+\eta^2|\ll 1$ in order
for \eqr{eq:AS} and \eqr{eq:AG} to be the correct second-order
expressions.  See Eq.\ (58) of \re{SL}.} We have therefore included
terms of order $\ep^2\xi$ and $\ep\eta\xi$, since they can contribute
to $1-n_S$ to second-order (i.e., they can have magnitudes similar to
those terms of order $\ep^2$ or $\ep\eta$).  This leads to the
additional terms $4\ep^2\xi C(C+1)$ and $-2\ep\xi\eta C^2$ in the
expression for $1-n_S$, which were not included in Ref.~\re{SL}.

Now that we have the complete expression to second order, it is easy
to isolate the first-order terms:\footnote{We include the $\eta/\ep$
and $\xi$ terms in the first-order expression for $1-n_S$ because they
{\it can} be greater than or of order one.}
\ba
\label{eq:allfirst}
R & = & \frac{25}{4}\ 2\epsilon \nonumber \\
n_T & = & -2\epsilon \nonumber \\
1-n_S & = & 2\epsilon \left[1 + \left(1-\frac{\eta}{\ep}+C\xi\right)\right].
\ea
The term corresponding to $\zeta$ in \eqr{eq:allfirst} is
$2\ep(1-\eta/\epsilon)$.  If $\eta=\ep$ and $|\xi|\ll1$, then
$1-n_S=2\epsilon=R/6.25$, and Eq.\ (\ref{eq:tsold}) is correct to
first order.  It is straightforward to show that $ \zeta \equiv
[2V'/(3H^2)]' = (6.25)^{-1}R(\eta/\epsilon-1)-2\ep\xi/3$.  Clearly in
terms of the slow-roll parameters, $\zeta$ can be of the same order as
$1-n_S$, $R$, or $n_T$ when $\eta\neq\ep$.  Eq.\ ({\ref{eq:allfirst}})
can now be rewritten as
\be
R=\frac{6.25\left(1-n_S+\zeta\right)}{1+\xi(C-1/3)}.
\ee
We can now see that
\eqr{eq:tsntold} holds only when $|\xi|\ll 1$.

The $\zeta$ term will contribute to the value of $1-n_S$ if $\eta\neq
\ep$ or if $|\xi|\ga |\eta/\ep|\ga 1$.  As an example, if $|\eta|/\ep\gg 1$ and
$|\xi|\ll 1$ (as is the case for natural inflation), then $1-n_S\simeq
-2\eta\simeq -R^{-1}dR/d \ln\lambda$, as we will see in Section III.
In this case, $1-n_S$ depends on the {\it derivative} of $R$ divided
by $R$, rather than on $R$.

Finally, note that the term proportional to $\xi$ in Eq.\
({\ref{eq:allfirst}}) comes from the derivative of a second-order term
in the expansion for $A_S$.  Thus, it cannot be derived from the
first-order result.  However, if $A_S$ is expanded to higher orders,
no higher-order derivative terms will appear.  This is because there
are no $\eta'$, $\eta''$, etc., terms in the expression for $A_S$
\re{SL}, since it was derived under that assumption that $\ep$ and
$\eta$ are constant as a perturbation evolves outside the Hubble
radius during inflation.

\vspace{36pt}
\centerline{\bf II. EXAMPLES}
\vspace{24pt}
\def\v{{V(\phi)}}

Let us now illustrate this formalism by a couple of examples.  Most of
these examples have already been worked out by the authors given in
the reference section.  They are given here not only to illustrate the
formalism developed in the last section, but also to make the case
that \eqr{eq:tsold} does not hold for a collection of popular
inflation models.

Only in a few cases, such as power-law inflation, can one find an
analytic solution for $H(\phi)$ and its derivatives.  It is more
useful to have an expression for the slow-roll parameters in terms of
the inflaton potential and its derivatives.  This is a difficult task
however, since the slow-roll parameters cannot be unambiguously
expressed in terms of the inflaton potential and its derivatives.
This is because $V^{(5)}/V'''$ and higher-order derivatives contribute
to the slow-roll parameter expressions, even though there is no
restriction on their magnitudes.  This point is elaborated in the
Appendix.

Starting with the field equations [\eqr{eq:eom}], we can express
$\epsilon$, $\eta$, and $\xi$ in terms of $V(\phi)$ and its
derivatives, appearing in various combinations involving
\be
\left(\frac{d^aV(\phi)}{d\phi^a}\right)^b
\left(\frac{d^cV(\phi)}{d\phi^c}\right)^{-d}; \qquad ab - cd=2.
\ee
To second order in $\{ \alpha,\ \beta,\ \gamma,\ \delta \} $ defined as
\be
\label{eq:albegade}
\kappa^2\alpha \equiv (V'/V)^2; \quad \kappa^2\beta \equiv V''/V; \quad
\kappa^2\gamma \equiv V'''/V';
\quad \kappa^2\delta\equiv V''''/V'',
\ee
the expressions for $\{\epsilon,\ \eta,\ \xi \}$ when $|\xi|\ll 1$
are\footnote{Although in this section we express results as an
expansion in the parameters $\alpha$, $\beta$, $\gam$, and $\del$, it
is important to remember that they are {\em not} the slow-roll
parameters, and the slow-roll parameters cannot be expressed
unambiguously in terms of them.}
\ba
\label{eq:epal}
\epsilon & = & \frac{1}{2}\alpha - \alpha\left[ \frac{\alpha}{3} -
\frac{\beta}{3}\right]  \nonumber \\
\eta & = &  - \frac{1}{2}\alpha + \beta +\alpha\left[\frac{2\alpha}{3} -
\frac{4\beta}{3} + \frac{\gamma}{3}  \right] +\beta
\left[\frac{\beta}{3}\right] \nonumber \\
\xi & = &  \frac{3}{2}\alpha - 3\beta + 2\gamma
-\alpha\left[ \frac{20\alpha}{3}-\frac{50\beta}{3}+\frac{13\gamma}{3} \right] -
\beta\left[ 9\beta - \frac{8\gamma}{3} - \frac{2\delta}{3} \right] .
\ea
These expressions are derived in the Appendix.  We can see that
$|\xi|\ll 1$ is satisfied when $|\gam|\ll 1$ and $|\bet\del|\ll 1$.
The more general solutions for $\ep$, $\eta$ and $\xi$ when $\gam$ and
$\del$ are larger than or are of order one are given also in the
Appendix.

Note that the first-order expressions for $\{\epsilon,\ \eta,\ \xi \}$
in the expansion in terms of $\{ \alpha,\, \beta,\, \gamma,\, \delta
\}$ can be found by ignoring all the terms in square brackets.

We can then substitute the above expressions into
\eqr{eq:TSgen}-({\ref{eq:nsgen}}) to obtain $R$ and the spectral
indices directly in terms of $\{ \alpha,\, \beta,\, \gamma,\, \delta
\}$.  To second order the expressions are
\ba
\label{eq:rntns}
R & = & \frac{25}{4} \alpha  + \frac{25}{4} \alpha \left[ \,
\frac{2}{3}(3C-1)(\alpha - \beta)
		\right] \nonumber \\
n_T & = & -\alpha -\alp \left[ \, \frac{11}{6}\alpha - \frac{4}{3}\beta +
2C(\alpha-\beta) \right]	\nonumber \\
1-n_S & = & 3\alpha-2\beta +\alpha \left[ \left( \frac{5}{6}+6C \right)\alpha +
(1-8C)\beta+ \left(2C-\frac{2}{3}\right)\gamma\right] -
\frac{2}{3}\beta\left[\beta\right] .
\ea
Again, the first order expressions can be obtained by setting the
terms in the square brackets to zero.  The first order expression when
$|\gamma|\ll 1$ is $1-n_S=3\alp-2\beta$, as was found first in Ref.\
\re{LiddLyth}.

It is shown in the Appendix that if $\gam>1$ and $\del$ is arbitrarily
large, the corrections to $R$ and $n_T$ are small, but the
corrections to the scalar spectral index, however,
can be large:
\be
1-n_S =
3\alp-2\bet+\left(2C-\frac{2}{3}\right)\alp\gam+\frac{2}{3}(C-1)\alp\bet\del.
\ee
Here derivatives of order $V^{(5)}/V'''$ and higher have been
neglected.  This new expression will be important for certain regimes
in Coleman-Weinberg, scale-invariant, and other models of inflation
for which $|\gam|\gg 1$ or $|\bet \del|\gg 1$.  Note that it is
possible for the $\alp\gam$ term to be of order the $\alp$ or $\bet$
terms, thus giving a different expression for $1-n_S$ to lowest order.
We are unable to find an inflaton potential for which this occurs,
however.

The procedure we will follow is straightforward.  For a given
potential, we calculate $\{ \alpha,\, \beta,\, \gamma,\, \delta \}$
using \eqr{eq:albegade} and check to see if $|\gam|\ll 1$ and
$|\bet\del|\ll 1$.  Then if we wish we can calculate the slow-roll
parameters using \eqr{eq:epal}, and then find $R$ and the spectral
indices using \eqr{eq:TSgen}-({\ref{eq:nsgen}}), or we can substitute
$\{ \alpha,\, \beta,\, \gamma,\, \delta \}$ directly into
\eqr{eq:rntns}.  We now turn to the examples.

\vspace{18pt}
\centerline{\bf A. Power-law inflation}

As a first example, we consider the second-order corrections for
power-law inflation, with potential $V(\phi)= M^4\exp(-2\phi/\phi_0)$.
For power-law inflation, it is simple to show that
$\alpha=\beta=\gamma=\delta=4/(\kappa^2\phi_0^2)$.  For power-law
inflation the second-order terms in $\{\epsilon,\ \eta,\ \xi \}$ all
vanish, and to second order $\epsilon = \eta = \xi =
2/(\kappa^2\phi_0^2)$. Therefore $|\xi|\ll 1$ and to second order
\ba
R & = & \frac{25}{2}\epsilon = \frac{25}{2}\frac{2}{\kappa^2\phi_0^2} \nonumber
\\
n_T & = & -2\epsilon[1+\epsilon] = -2 \frac{2}{\kappa^2\phi_0^2} \left[ 1 +
\frac{2}{\kappa^2\phi_0^2} \right] \nonumber \\
1-n_S & = & 2 \epsilon [1 + \epsilon] = 2 \frac{2}{\kappa^2\phi_0^2} \left[1+
\frac{2}{\kappa^2\phi_0^2} \right].
\ea
So the relationship between $R$, $n_T$, and $1-n_S$ to second order is
\be
-6.25n_T = 6.25(1-n_S) = R\left[1+\frac{2}{25}R\right].
\ee
For the exponential potential the second-order corrections are of
order $8R\%$ of the first-order term.  The magnitude of the
second-order corrections increase with $R$, and can become important.
We can also express the second-order corrections in terms of the
parameters of the potential by writing $2R/25 = 2/(\kappa^2\phi_0^2) =
\mpl^2/(4\pi\phi_0^2)$.

Only for the exponential potential of power-law inflation is
$\epsilon=\eta$.  This results in $\zeta\simeq 0$.

These results for power-law inflation are summarized in Table 1.

\vspace{18pt}
\centerline{\bf B. Chaotic and hybrid inflation}

Now let's consider a potential commonly used in chaotic inflation \re{Chao}:
\be
V=v \phi^p,
\ee
where $v$ and $p$ are constants.  The mass dimension of $v$ is $4-p$,
and $p$ is an integer.  We denote as $\phi_N$ the value of $\phi$
corresponding to the value of the scalar field when the length scale
of interest crossed outside the Hubble radius during inflation.  The
expansion parameters $\{ \alpha,\ \beta,\ \gamma,\ \delta \}$ are
easily found, giving to second order
\ba
\epsilon & = & \frac{1}{2\kappa^2\phi_N^2}\left[ p^2
- \frac{2p^3}{3\kappa^2\phi_N^2} \right]	 	    \nonumber \\
\eta & = & \frac{1}{2\kappa^2\phi_N^2}\left[ p(p-2)
- \frac{2p^2(p-3)}{3\kappa^2\phi_N^2} \right]               \nonumber \\
\xi & = & \frac{1}{2\kappa^2\phi_N^2} \left[ (p-2)(p-4)
- \frac{2p(p^2-9p+28)}{3\kappa^2\phi_N^2} \right] .
\ea
Since $\eta\simeq\xi$, the condition $|\xi|\ll 1$ is satisfied.  It
will be convenient to define the dimensionless ratio $r_N =
1/(2\kappa^2\phi_N^2) = \mpl^2/(16\pi\phi_N^2)$.  The ratio $r_N$ can
be related to the number of $e$-folds to the end of inflation:
$r_N=1/(4Np)$.  For $p=4$, $r_N\simeq 1/(16N)$ and for $p=2$,
$r_N=1/(8N)$. Since for the scales of interest to us $N\sim 50$,
$r_N\sim 1.25\times10^{-3}$ for $p=4$ and $r_N\sim 2.5\times10^{-3}$
for $p=2$.  The expressions to second order are
\ba
R & = & \frac{25}{4} \, 2r_Np^2\left[ 1 - \frac{4}{3}r_N p(1-3C)\right]
\nonumber \\
n_T & = & -2r_Np^2 \left[ 1+ r_Np^2 + \frac{4}{3}r_Np(2+3C)\right] \nonumber \\
1-n_S & = & 2r_Np^2\left[ 1 + \frac{2}{p} + \frac{1}{3}r_N(3p^2+p(14+12C)
-12(1-2C)) \right].
\ea
Note that our expansion is valid in this example, because $|\xi|\la
r_N\ll 1$ for $p\leq 4$.  For $p=2$, the terms in the square brackets
for $R$, $n_T$ and $1-n_S$ are $[1-8.5r_N]$, $[1+3.5r_N]$, and
$[2-2.3r_N]$ respectively, while for $p=4$ the terms are $[1-17r_N]$,
$[1+15r_N]$ and $[1.5 +13r_N]$.  Therefore the largest second-order
corrections is $17 r_N\sim 2.1 \%$ for $p=4$.

Now turn to the first-order results. To first order, the relationship
between $R$, $n_T$, and $1-n_S$ is
\be
R = -6.25n_T = 6.25(1-n_S)/(1+2/p).
\ee
If we take $p=2$, $4$ and $\infty$, then $R\simeq 3.1(1-n_S)$,
$4.2(1-n_S)$ and $6.25(1-n_S)$, respectively.  Thus, the contribution
of the $\zeta$ term depends upon $p$, and is negligible only for $p\gg
1$.

Note that the tensor to scalar ratio for this model need not be {\em
very} small, since $R=(25/2)r_Np^2$, which for $p=2$ and $4$ is
$R=0.125$ and $0.25$.  In any case, in slightly more complicated
inflation models it is possible to have inflation during an epoch when
the potential is approximately power law, but to modify the relation
between $r_N$ and $N$.

As an example of such a modified model, we examine a hybrid
inflationary model [{\ref{Hyb}}] of two scalar fields with potential
\ba
V(\sig,\phi) & = & \frac{1}{4\lam} \left( M^2-\lam\sig^2 \right)^2 +
\frac{m^2}{2}\phi^2 +\frac{g^2}{2}\phi^2\sig^2.
\ea
When $\phi>\phi_C \equiv M/g$, the global minimum is at $\sigma=0$.
Assuming that $\sig\simeq 0$, the effective potential during inflation
is
\ba
V(\phi)&\simeq&\frac{M^4}{4\lam}+\frac{m^2}{2}\phi^2.
\ea
Except for the addition of a constant term, the potential looks like
the $p=2$ version of chaotic inflation.

We will assume that $M^2> 2\lam m^2/g^2$.  During the first epoch of
inflation $V\simeq m^2\phi^2/2$, and during the second epoch $V\simeq
M^4/(4\lam)$. The value of the field at the transition between these
two epochs is given by $\phi_T =M^2/(\sqrt{2\lam}m)$. We will also
assume that $M^3\ll\sqrt{\lam}gm ~m_{Pl}^2$ (or $2\phi_T ^2\phi_C
^2\ll m_{Pl}^4$) so that inflation ends immediately after $\phi\simeq
\phi_C$ \re{Hyb}.

In order to satisfy the slow-roll conditions during inflation,
$\phi_T/m_{Pl}\gg 1/\sqrt{12\pi}$.  Assuming that $\sig\simeq 0$, the
slow-roll parameters are
\ba
\kappa^2\ep & = & \frac{2\phi_N^2}{(\phi_N ^2+\phi_T^2)^2}
\left[ 1 - \frac{4}{3}\frac{\phi_N^2-\phi_T^2}{\kappa^2(\phi_N^2+\phi_T^2)^2}
\right]   \nonumber \\
\kappa^2\eta & = &  \frac{2\phi_T ^2}{(\phi_N ^2 +\phi_T^2)^2}
\left[ 1 + \frac{2}{3}\frac{1}{\kappa^2\phi_T^2}
- \frac{16}{3}\frac{\phi_N^2}{\kappa^2(\phi_N^2+\phi_T^2)^2} \right] \nonumber
\\
\kappa^2\xi & = & - \frac{6\phi_T ^2}{(\phi_N ^2 +\phi_T^2)^2}
\left[ 1 + \frac{6}{\kappa^2\phi_T^2} - \frac{40}{9} \frac{\phi_N^2}{\phi_T^2}
\frac{(\phi_N^2+5\phi_T^2)}{\kappa^2(\phi_N^2+\phi_T^2)^2} \right] .
\ea

Note that for $\phi_N\gg\phi_T$ the results for $p=2$ chaotic
inflations obtains. We will be interested in the opposite limit,
$\phi_N\ll\phi_T$.  In this case
\ba
 \ep & = & \frac{\phi_N^2}{\phi_T^2} \frac{2}{\kappa^2\phi_T^2}
\left[ 1 + \frac{4}{3\kappa^2\phi_T^2} \right]   \nonumber \\
\eta & = &  \frac{2}{\kappa^2\phi_T^2}
\left[ 1 + \frac{2}{3}\frac{1}{\kappa^2\phi_T^2} \right]
\nonumber \\
\xi & = & - \frac{6}{\kappa^2\phi_T^2}
\left[ 1 + \frac{6}{\kappa^2\phi_T^2} \right] .
\ea

In this case $|\ep| \ll \{|\eta|,\ |\xi|\}$ but $|\xi|\ll 1$ since
$\eta\simeq -\xi/3$.  The expressions to second order are
\ba
R&=&\frac{25~\phi_N^2}{\kappa^2\phi_T^4}
\left[1+\frac{4(1-3C)}{3\kappa^2\phi_T^2}\right]\nonumber\\
n_T&=&-\frac{4~\phi_N^2}{\kappa^2\phi_T^4}
\left[1-\frac{16(2+3C)}{3\kappa^2\phi_T^2}\right]\nonumber\\
1-n_S&=&-\frac{4}{\kappa^2\phi_T^4}\left[1+\frac{2}{3\kappa^2\phi_T^2}\right].
\ea
To illustrate the point that it is possible to have $R\sim0$ with
$1-n_S$ relatively large, we can work to first order in $\{\epsilon,\
\eta,\ \xi \}$ and first order in $\phi_N/\phi_T$.  In these limits
\be
R \sim n_T \sim 0; \qquad 1-n_S \sim -2\eta  \sim -8r_T,
\ee
where we have defined $r_T=1/(2\kappa^2\phi_T^2)$ in the same manner
as we have defined $r_N$, so that the slow-roll condition is satisfied
for $r_T\ll 3/4$).  Note that $n_S>1$ in this case, because
$|\eta/\ep|\gg 1$, and because the second derivative of the inflaton
potential is positive: $V''>0$.

The number of {\em e}-folds from the end of inflation is
\be
\label{eq:Nhyb}
N=-\frac{8\pi}{m_{Pl}^2}\int\frac{Vd\phi}{V'}
=\frac{1}{8r_N}\left[1+\frac{r_N}{r_T }\ln(r_C /r_N)-\frac{r_N}{r_C }\right],
\ee
where $r_C$ is defined to be $r_C \equiv 1/(2k^2\phi_C ^2)>r_T $.
Since we are interested in $N\sim 60$, then $r_C \gg r_N$.  In
addition, we normalize to the COBE results by
calculating the density
fluctuation amplitude,
\be
\delta_H=\frac{4\sqrt{2V}}{5\sqrt{3} ~m_{Pl}^2 \sqrt{\ep}},
\ee
where $V$ and $\ep$ are evaluated when the scale $\lambda$ left
the Hubble radius.
If $\delta_H$ is evaluated $60$ {\em e}-folds before the end of inflation,
then $\delta_H\simeq 1.7\times 10^{-5} $\re{CLLSW}.
It is shown in Ref. \re{CLLSW} that there is only
a small region in parameter space allowed after imposing
this constraint.

We first choose an example whereby the $\zeta$-term completely
dominates, but for which $R$ is very small.  We take $M=9\times
10^{-4}m_{Pl}$, $m=5.4\times 10^{-7}m_{Pl}$ and $\lam=g=1$.  For this
model, the number of {\em e}-folds in the second inflationary epoch is
$100$.  Therefore, the observable universe would have density
perturbations only from the constant potential epoch for reasonable
reheat temperatures.  For this model, $r_T\simeq 0.0177$.  For the
scale leaving 60 {\em e}-folds before the end of inflation,
$R=3.0\times 10^{-3}$, $n_T=-4.9\times 10^{-4}$ and $n_S=1.14$ to
first order.  The value for $R$ ignoring $\zeta$ is
$6.25(1-n_S)=-0.88$, which is not only negative (and therefore is not
physically meaningful) but is also $294$ times too large.  Clearly
$R\neq 6.25(1-n_S)$ because the $\zeta$ term overwhelmingly dominates:
$\eta/\ep=r_N/r_T=286$.  We can also find a value for $m$ such that
$R\sim 1$ but the $\zeta$-term still contributes non-negligibly.  We
take $M=2.5\times 10^{-3}m_{Pl}$, $m=4.857 \times 10^{-6}m_{Pl}$ and
$\lam=g=1$.  For this model, the number of {\em e}-folds in the second
inflationary epoch is $61$ and $r_T=0.024$.  In this region in
parameter space, the first-order contributions vanish because
$\eta/\ep \simeq 2$ when $N_{\phi}=60$.  For a scale leaving 60 {\em
e}-folds before the end of inflation, $R=0.26$ and $n_T=-0.042$ to
first order, and $1-n_S=1.8\times 10^{-3}$ to second-order.  Thus the
value for $R$ ignoring $\zeta$ is $6.25(1-n_S)=0.011$, which is a
factor of $24$ times too small.  We see that even when $R$ is near
one, the ``correction'' terms can contribute nearly $100$\%.

\vspace{18pt}

\centerline{\bf C. Natural inflation}

In natural inflation \re{Nat} the potential is $V(\phi) = \Lambda^4 \left[ 1
+\cos(\phi/f) \right]$, with $f\sim\mpl\gg\Lambda$.  Inflation occurs when
$|\phi|\ll f$.  The spectral indices and $R$ are easily found to
second order in $\{ \alpha,\ \beta,\ \gamma \}$.  Since the
expressions are unwieldy, we expand the trigonometric functions in
$\phi/f$. To lowest order in $\phi/f$ but second order in $\{ \alpha,\
\beta,\ \gamma \}$, the expressions for $R$ and the spectral indices are
\ba
R & = & \frac{25}{4} \left( 1 + \frac{3C-1}{3\kappa^2f^2} \right)
   \frac{1}{4\kappa^2 f^2}\left( \frac{\phi_N^2}{f^2} \right)\nonumber \\
n_T & = & -\left( 1 + \frac{3C+2}{3\kappa^2f^2} \right)
\frac{1}{4\kappa^2 f^2}\left( \frac{\phi_N^2}{f^2} \right) \nonumber \\
1-n_S & = & \left( 1 -\frac{1}{6\kappa^2f^2} \right)\frac{1}{\kappa^2 f^2}
\ea
where again $\phi_N$ is the value of $\phi$ for the length scale of
interest.  Note that $2\bet\simeq\gam\simeq\del$, so that $|\xi|\ll
1$.  It is clear that since $|\phi_N|/f\ll1$, $R\sim -n_T \ll1$, but
$1-n_S$ can be substantial.

Now let's consider the $\zeta$ contribution in the first-order result.   To
first order
\be
\epsilon= \frac{1}{8\kappa^2f^2} \left( \frac{\phi_N^2}{f^2} \right)
\left(1-\frac{1}{3\kappa^2 f^2}\right) ;
\qquad \eta = - \frac{1}{2\kappa^2f^2}\left(1-\frac{1}{6\kappa^2 f^2}\right) ,
\ee
so $\eta|/\epsilon\gg 1$ is large, and in fact the $\zeta$
contribution, proportional to $1-\eta/\epsilon$, dominates. Thus, to
first order, $1-n_S$ is {\em independent} of $R$, although
$-6.25n_T\sim R$ is still valid. We remind the reader that for this
model $R\ll1$.

\vspace{18pt}

\centerline{\bf D. Scale-invariant inflation}
\def\phit{{{\widetilde{\phi}}}}

We now consider the potential found in a scale-invariant theory
[{\ref{Scale}}]:
\be
V(\phi)=\Lambda^4\left[1+\frac{\phi-\phit}{\phit}\exp(\phi/\phit)\right],
\ee
where $\Lambda$ and $\phit>0$ are positive constants with mass
dimension $4$ and $1$, respectively.  This potential has a global
minimum at $\phi=0$, and slow-roll inflation occurs for $\phi=\phi_N$
when $|\phi_N/\phit|\gg 1$ in the regions $\phi_N> 0$ or $\phi_N< 0$.
In the region of positive $\phi$, the results resemble power-law
inflation:
\be
\frac{4}{25}R = -n_T = 1-n_S = \frac{1}{\kappa^2\phit^2};
\qquad \phi/\phit \gg1
\ee
Note that the results do not depend upon $\phi_N$ (to leading order).  Thus
$R$, $n_T$, and $n_S$ are truly constant.

The results in the other region of inflation ($\phi/\phit\ll-1$) are more
interesting:
\ba
\label{eq:SIstuff}
R & = & \frac{25}{4} \frac{1}{\kappa^2\phit^2} \left( \frac{\phi_N}{\phit}
\right)^2 \exp(-2|\phi|/\phit) \nonumber \\
n_T & = & -\frac{1}{\kappa^2\phit^2} \left( \frac{\phi_N}{\phit} \right)^2
\exp(-2|\phi|/\phit) \nonumber \\
1-n_S & = & \frac{2}{\kappa^2\phit^2} \left( \frac{|\phi_N|}{\phit} \right)
\exp(-|\phi|/\phit).
\ea
Because $\gam\simeq\del=1/(\kappa^2\phit^2)$, then $\kappa^2\phit^2\gg
1$ in order that \eqr{eq:SIstuff} is valid.  The second-order
corrections are $1/(\kappa^2\phit^2)$ for $1-n_S$ and
$-1/(\kappa^2\phit^2)|\phi_N|/\phit\exp(-|\phi_N|/\phit)$ for $R$ and
$n_T$.  In this case $R\sim -6.25n_T$, but $1-n_S \propto \sqrt{R}$.
In addition, note that $R$ is small. One can imagine $1-n_S$ large
enough to be detectable, but $R\sim 0$.

Because $|\alp\gam|\ll \bet$, when $\gam>1$ the scalar index is
$1-n_S\simeq -2\bet$ to first order, as given above in \eqr{eq:SIstuff}.

\vspace{18pt}

\centerline{\bf E. Coleman-Weinberg inflation}


Finally, we examine the Coleman-Weinberg potential in the context of
new inflation.  The potential is
\be
V(\phi)=B\sig^4/{2}+B\phi^4[\ln(\phi^2/\sig^2)-1/2].
\ee
The global minimum of this potential is at $\phi=\sig$.  Inflation
occurs for $0<\phi/\sig\ll 1$ when the potential is nearly flat:
$V\simeq B\sig^4/2$.  We can calculate the slow-roll parameters in
this model with the assumption that $V\sim B\sig^4/{2}$ in the
denominator of $\alpha$ and $\beta$.  The spectral indices and $R$ are
to lowest order in $\phi_N/\sigma$
\ba
R & = & -\frac{25}{4}n_T= \frac{25}{4}\frac{64}{\kappa^2\sigma^2}
\left(\frac{\phi_N}{\sig}\right)^6
\left[\ln\left(\phi_N^2/\sig^2\right)\right]^2  \\
\label{eq:nSCW}
1-n_S & = &  \frac{48}{\kappa^2\sigma^2}
\left(\frac{\phi_N}{\sig}\right)^2
|\ln\left(\phi_N^2/\sig^2\right)| =\frac{3}{N_{\phi}}.
\ea
Note that $\gam=3\del=6/(\kappa^2\phi_N^2)$ in this limit, so that
$\kappa^2\phi_N^2\gg 6$ in order for these expressions to be valid.
To first order then and for any value of $\sig < 3 m_{Pl}$,
$1-n_S\propto const$ for a constant value of $N_{\phi}$.

Again, the fact that $n_T \neq n_S-1$ can be traced to a large value
of $\zeta$, i.e., $|\eta/\epsilon| \neq 1$.  Let's look at the
slow-roll parameters with the assumption that $\phi_N/\sig\ll 1$:
\ba
\kappa^2\ep&=&\frac{32}{\sigma^2}\left(\frac{\phi_N}{\sig}\right)^6
\left[\ln\left(\phi_N^2/\sig^2\right)\right]^2; \quad
\kappa^2\eta = \frac{24}{\sigma^2}\left(\frac{\phi_N}{\sig}\right)^2
\ln\left(\phi_N^2/\sig^2\right) \nonumber \\
\kappa^2\xi&=&\frac{12}{\sigma^2}\left(\frac{\sig}{\phi_N}\right)^2.
\ea
Note that $|\xi| \gg |\eta| \gg |\epsilon|$, so that the $\zeta$
contribution is dominant: $\eta/\ep\simeq (3/4)(\sig/\phi_N)^4/$
$\ln(\phi_N/\sig)^2 \ll -1$.  Therefore, for $1-n_S$ the second-order
contribution will be of order $3/(2\pi)~(m_{Pl}/\phi_N)^2$, and
$(3/\pi)~(m_{Pl}^2\phi_N^2/\sig^4)\ln(\phi_N^2/\sig^2)$ for $R$ and
$n_T$.

Because $|\alp\gam|\ll \bet$, when $\gam>1$ the scalar index is
$1-n_S\simeq -2\bet$ to first order, as given above in \eqr{eq:nSCW}.

Although $R$ is very small for most values of $\sig$, if $\sig$ is
large enough, $R$ can increase to as much as $8$\%.  We can then
numerically solve the exact expressions.  Setting $N_{\phi}=50$ and $\sig=10
m_{Pl}$, we find that $R=0.083$ and $n_S=0.962$.  Thus, if we
neglect the $\zeta$ term, $6.25(1-n_S)=0.24$ which is almost $3$ times
larger than the correct value.

A summary of the results of this section is given in Table 1.

\renewcommand{\arraystretch}{1.3}
\begin{table}[t]
\caption{Magnitude of corrections to spectral relations when $|\xi|\ll 1$
($x\equiv
\phi_N/\phit$).}
\begin{center}\begin{tabular}{lcc}
& & \\\hline\hline
Inflation & \phantom{XX} Second-order \phantom{XX} & Relative contribution of
\\
Model     & correction   & $\zeta$ term: $|1-\eta/\epsilon|$ \\ \hline
Power-law & $8R\%$ & 0 \\
Chaotic ($p\leq 4$) & $\leq 2.1\%$ & $2/p$: $100\%~(p=2);~~50\%~(p=4)$ \\
Hybrid ($r_N\ll r_T$) & $\leq 2.1\%$ & $100\%$ \\
Hybrid ($r_N\gg r_T$) & $r_T$  & $(\phi_T/\phi_N)^2\gg 1$ \\
Natural & $\mpl^2/(16\pi f^2)$ & $4f^2/\phi_N^2\gg1$ \\
Scale-Invariant ($x\gg1$)& $1/(\kappa^2\phit^2)$ & $0$\\
Scale-Invariant ($x\ll-1$) & $\la 1/(\kappa^2\phit^2)$ &
$2|x|^{-1}e^{|x|}\gg1$\\
Coleman-Weinberg & $ \la (3/(2\pi))(m_{Pl}/\phi_N)^2$ &
$(3/4)(\sig/\phi_N)^4/|\ln(\phi_N^2/\sig^2)|\gg1$ \\
\hline\hline
 & & \\
 & & \\
\end{tabular}\end{center}
\end{table}

\vspace{36pt}
\centerline{\bf III. First-Order Expression relating $n_S$, $R$ and derivatives
of $R$}

\vspace{24pt}

Using Eqs.\ ({\ref{eq:ep}})-({\ref{eq:xi}}) and
Eq.\ ({\ref{eq:dlamdphi}}), we can express $\eta$ and $\xi$ in terms of
$\ep$ only.
\ba
\label{eq:etaeplam}
\eta&=&\ep+\frac{1-\ep}{2\ep}\frac{d\ep}{d\ln\lambda}\\
\xi&=&\eta+\frac{1-\ep}{\ep}\frac{d\eta}{d\ln\lambda}\\
 &=& \eta+\frac{1-\ep}{\ep}
\left( \frac{d\ep}{d\ln\lambda} -
\frac{1}{2\ep^2}\left(\frac{d\ep}{d\ln\lambda} \right)^2
+ \frac{1-\ep}{2\ep}\frac{d^2\ep}{d\ln\lambda^2} \right).
\ea

Using Eq.\ ({\ref{eq:allfirst}), we then find that
the most general first-order expression relating
$n_S$, $R$ and derivatives of $R$ with respect to $\ln\lam$ is
\be
\label{eq:nSnTgen}
6.25(1-n_S)\simeq R\left[1-\frac{6.25}{R^{2}}
\left( \frac{dR}{d\ln\lambda} -C\frac{d^2R}{d\ln\lambda^2}\right)
\right].
\ee
This shows explicitly when the old formula fails.  If
$R^{-2}|dR/d\ln\lam-C d^2R/d\ln\lam^2|\gg 1$, then the ``correction
term'' dominates and $6.25(1-n_S)\neq R$.  Thus even if $R(\lambda)$
changes slowly changes with scale, if $R\ll 1$, ``corrections'' to the
previous formula can dominate.  Note that the $\eta/\ep-1$ and $\xi$
terms are the $dR/d\ln\lam$ and $d^2R/d\ln\lambda^2$ terms,
respectively.

\vspace{36pt}
\centerline{\bf IV. CONCLUSIONS}
\vspace{24pt}

In this paper, we derive the contribution of scalar and tensor
perturbations from inflation to second order in slow-roll parameters.
We find that the previously derived formula fails when $\eta\neq\ep$
or $|\xi|\ga 1$.  In particular, it fails for natural inflation and
Coleman-Weinberg inflation, where $|\eta|\gg\ep$, and for ``chaotic''
$\phi^2$ inflation, where $|\eta|\ll \ep$. For natural inflation, a
type of scale-invariant inflation, and Coleman-Weinberg inflation, to
first order $1-n_S\simeq const$, $1-n_S\propto \sqrt{R}$ and
$1-n_S\propto const$, respectively.  Thus the relationship between $R$
and $1-n_S$ is in general not linear.  We have shown that this occurs
when $V|V''|/(V')^2\neq 1$ or $m_{Pl}^2/(4\pi) ~|V'''/V'|\ga 1$.
We have also found that the slow-roll parameters, the ratio
of the tensor to the scalar amplitude,
and the scalar and tensor indices cannot be
expressed unambiguously in terms of the potential and its derivatives
unless the higher-order derivatives of the potential are small.
In addition, we calculated the general expression for $\xi$ when $\xi>1$.
This alters the expression for the scalar spectral index
when higher-order derivative terms are marginally large.

\vspace{18pt}

\centerline{\bf ACKNOWLEDGMENTS}

EWK is supported by the DOE and NASA under Grant NAGW--2381.
SLV was supported by the President's Postdoctoral Fellowship Program
at the University of California and NSF Grant AST-9120005.
We would like to thank S.\ Arendt, A.\ R.\ Liddle, and M.\ S.\ Turner
for useful discussions.

\frenchspacing
\def\prl#1#2#3{Phys. Rev. Lett. {\bf #1}, #2 (#3)}
\def\prd#1#2#3{Phys. Rev. D {\bf #1}, #2 (#3)}
\def\plb#1#2#3{Phys. Lett. {\bf #1B}, #2 (#3)}
\def\npb#1#2#3{Nucl. Phys. {\bf B#1}, #2 (#3)}
\def\apj#1#2#3{Astrophys. J. {\bf #1}, #2 (#3)}
\def\apjl#1#2#3{Astrophys. J. Lett. {\bf #1}, #2 (#3)}
\begin{picture}(400,50)(0,0)
\put (50,0){\line(350,0){300}}
\end{picture}

\vspace{0.25in}

\def\labelenumi{[\theenumi]}

\begin{enumerate}
\item\label{BUNCH} A.A. Starobinskii, {\em Sov. Astron. Lett.}, {\bf 11},
	133 (1985); L. M. Krauss and M. White, \prl{69}{869}{1992};
  	J. E. Lidsey and P. Coles, {\em Mon. Not. Roy. astr. Soc.} {\bf 258},
  	57P (1992); D. S. Salopek, \prl{69}{3602}{1992}; D. S. Salopek,
  	in {\em Proceedings of the International School of Astrophysics
  	``D. Chalogne'' second course}, ed N. Sanchez (World Scientific,
	1992); T. Souradeep and V. Sahni, {\em Mod. Phys. Lett.} {\bf A7},
	3541 (1992); M. White, \prd{46}{4198}{1992}; R. Crittenden, J. R.
	Bond, R. L. Davis, G. Efstathiou and P. J. Steinhardt, in {\em
	Proceedings of the Texas/Pascos Symposium, Berkeley} (1992).

\item\label{CKLL}  E. J. Copeland, E. W. Kolb, A. R. Liddle and J. E.
	Lidsey, \prl{71}{219}{1993};  \prd{48}{2529}{1993};
	M. Turner, {\em Phys. Rev D},{\bf 48}, 5539, (1993).

\item\label{CBDES} R. Crittenden, J. Bond, R. Davis, G. Efstathiou,
	P. Steinhardt, \prl{71}{324}{1993}; R. Davis,
	``Gravitational Waves and the Cosmic Microwave Background,''
	from the Proceedings of the 16$^{th}$ Texas/Pascos Symposium
	(Berkeley), 1993 ;J. Bond, R. Crittenden, R. Davis, G. Efstathiou
	and P. Steinhardt, \prl{72}{13}{1994}; R. Crittenden, R. Davis,
	P. Steinhardt, {\em Astroph. J}, {\bf 417}, L13 (1993).

\item \label{HJ} D. S. Salopek and J. R. Bond, \prd{42}{3936}{1990};
	J. E. Lidsey, \plb{273}{42}{1991}.

\item\label{LiddLyth} A. Liddle and D. Lyth, \plb{291}{391}{1992}.

\item \label{SL} E. D. Stewart and D. H. Lyth, \plb{302}{171}{1993}.

\item\label{CKLL3} E. J. Copeland, E. W. Kolb, A. R. Liddle and J. E.
	Lidsey, \prd{49}{1840}{1994}; A. Liddle and M. Turner,
	``Second-order Reconstruction of the Inflationary Potential'',
	Fermilab report FNAL--PUB--93/399--A (1993) and SUSSEX-AST 94/2-1.

\item\label{Chao} A. Linde, {\it Phys. Lett} {\bf 129B},177 (1983).

\item\label{Hyb} A. Linde, \prd{49}{748}{1994};
	A. Linde, ``Comments on Inflationary Cosmology'', SU-ITP-93-27, 1993.

\item\label{CLLSW} E. Copeland, A. Liddle, D. Lyth, E. Stewart
	and D. Wands, ``False Vacuum Inflation with Einstein Gravity'',
	to appear in {\em Phys. Rev.} {\bf D}.

\item\label{Nat} K. Freese, J.Frieman and A. Olinto, {\em Phys Rev Lett},
{\bf 65}, 3233,(1990).

\item\label{Scale}R. Holman, E. Kolb, S. Vadas and Y. Wang, {\em Phys. Rev D},
{\bf 43}, 3833, (1991); R. Holman, E. Kolb, S. Vadas and Y. Wang, {\it Phys.
Lett.}, {\bf 269B}, 252, (1991),

\end{enumerate}

\newpage
\nonfrenchspacing

\centerline{\bf{Appendix}}
\def\ksection{\arabic{section}}
\def\theequation{A.\arabic{equation}}
\setcounter{equation}{0}

In this appendix, we derive the first-order results
for $\ep$, $\eta$ and $\xi$ when $|\xi|\ga 1$, and the
second-order results when $|\xi|\ll 1$.
The latter example is used in Section II of this paper.

We define the function $f\equiv {H'}/{H}$.
Then using \eqr{eq:eom}, which can be rewritten as
\be
H^2=\frac{\kappa^2\,V}{3}\left(1+\frac{\ep}{3} \right),
\ee
and $\ep'=2f(\eta-\ep)$, ``$f$'' becomes
\be
f=\frac{1}{2}~\frac{V'}{V}\left[ \, 1 + \frac{\eta-\ep}{3} \,\right].
\ee
The slow-roll
parameters can then be determined in terms of $f$ and its derivatives:
$\ep=2f^2/\kappa^2$,  $\eta=2(f^2+f')/\kappa^2$
and $\xi=2(f^2+3f'+f''/f)/\kappa^2$.
In addition, the derivatives of $\eta$ and $\znew$
are $\eta'=f(\xi-\eta)$ and
$\xi'=f(\eta/\ep)~(\znew-\xi)$, where
\be
\znew\equiv \frac{2}{\kappa^2} \frac{H''''}{H''}.
\ee
The mixed second-order expressions for $\ep$, $\eta$
and $\xi$ as a function of $\znew$ and
$\{ \alpha,\ \beta,\ \gamma \} $ (as defined
in \eqr{eq:albegade}) are
\ba
\label{eq:epinit}
\ep&=&\frac{1}{2}\alp\left(1+\frac{2}{3}
\left\{\eta-\ep\right\} \right) \\
\label{eq:etainit}
\eta&=&
\bet\left(1+\left\{\frac{\eta-\ep}{3}\right\} \right)
-\frac{1}{2}\alp\left[\, 1-\frac{1}{3}\xi
+\left\{\eta-\frac{2}{3}\ep+\frac{\xi}{9}(\ep-\eta)\right\} \,\right]  \\
\label{eq:xiinit}
\xi&=&
2\,{\it \gam} - 3\,{\it \bet}\,\left[ 1 - {{\xi }\over 3} +
\left\{\frac{2\eta-\ep}{3}\right\} \right]  +
  {{3\,{\it \alp}}\over 2}\,\left[ 1 +\xi
            \left( -\frac{2}{3} -\frac{1}{9}\frac{\eta }{\ep} +
{{\xi }\over {27}} \right) + \frac{1}{9}\znew\frac{\eta}{\ep}
\right.\nonumber\\
& & \left.+\left\{ \frac{2}{3}(2\eta-\ep)+
\frac{\xi}{27}\left(4\ep
-5\eta-\eta\frac{\eta}{\ep}\right)
+\frac{\znew\,\eta}{27}\left(-1+\frac{\eta}{\ep}\right) \right\}
\right].
\ea
In the above equations, the first-order expressions can be obtained
by setting the terms in curly brackets $\{\}$ equal to zero.
Note that we cannot solve for $\xi$ (and therefore $\eta$ and $\ep$)
until $\znew$ is determined.

We calculate $\znew$ from $f$ and its derivatives:
\be
\label{eq:kz}
\frac{\kappa^2\znew}{2}=f^2+3f'+\frac{f''}{f}+\frac{f}{f^2+f'}
\left[\, 2ff'+3f''+\frac{f'''}{f}+\frac{f f''}{f^2}\,\right].
\ee
Because the expression for $\znew$ to second-order
is too long and lends no new insight,
we calculate $\znew$ to
first-order only.  We note that the second-order corrections
are of order $\ep$, $\bet$ and $\alp\xi$.
As is similar to the mixed expressions for $\ep$, $\eta$ and $\xi$
in \eqr{eq:epinit}-({\ref{eq:xiinit}}),
several terms on the right-hand side of \eqr{eq:kz} will contain the factors
$\znew$ and $\znew'$.  We substitute in
the result $\znew'=f~(\xi/\eta)(\tau-\znew)$,
where $\tau=(2/\kappa^2) H^{(5)}/H'''$.
Then, combining
all of the $\znew$ terms and keeping only the largest ones, we find that
\ba
\label{eq:zetinit}
\znew&=&
 \frac{2}{\eta}
\left( 1 + \frac{ \alp^2\xi}{36\eta} \right)^{-1}
\left[\,\alp\gamma\left[-4+\xi\right]
+2\beta\gamma
+ \beta\delta + \frac{\alp\xi~ \tau}{12} +\right.\nonumber\\
& &\left.  \alp^2\left[-\frac{47}{8}+\frac{\xi}{216}
\left(873+126\frac{\eta}{\ep}
+27\left(\frac{\eta}{\ep}\right)^2
-9\frac{\xi}{\ep}
-24~\xi\frac{\eta}{\ep}+3\xi^2\right)\right].
\right.\nonumber\\
& & \left.
+\alp\beta\left[\frac{29}{2}+\frac{\xi}{216}
\left(-1404-144\frac{\eta}{\ep}+84\xi\right)\,\right]
+\beta^2\left[-\frac{15}{2}+\frac{3}{2}\xi\,\right] \right],
\ea
The quantity $\tau$ is the higher order derivative term mentioned
at the beginning of Section II, and is of order $\kappa^{-2} V^{(5)}/V'''$
as long as higher order derivatives are unimportant.

As an example, if $|\xi|\ll 1$,
\ba
\label{eq:znewll}
\znew&=&
 \frac{2}{\eta}
\left[-4\alp\gamma +2\beta\gamma + \beta\delta
-\frac{47}{8}\alp^2+\frac{29}{2}\alp\beta
-\frac{15}{2}\beta^2\right].
\ea
Note that $\eta\znew$ is  second order in small quantities, so that
the ``$\alp\eta\znew/\ep$'' term
in \eqr{eq:xiinit} is second order.
We now calculate $\ep$, $\eta$ and $\xi$ in this limit
from \eqr{eq:epinit} - ({\ref{eq:xiinit}}).
For the second-order terms, we substitute in \eqr{eq:znewll}
and the first-order
expressions for $\ep$, $\eta$ and $\xi$, which are
$\ep=\alp/2$, $\eta= \bet-\alp/2$
and $\xi= 2\gam-3\bet+3\alp/2$.
The final results
(when $|\xi|\ll 1$) are given in \eqr{eq:epal}.

To determine the slow-roll parameter
$\xi$ when $1\la |\xi|\ll 1/\ep$, we substitute
\eqr{eq:zetinit} into \eqr{eq:xiinit} along
with the first-order expressions
\ba
\label{eq:epeta}
\ep&=&\frac{\alp}{2}\nonumber\\
\eta&=&\bet-\frac{\alp}{2}\left(1-\frac{\xi}{3}\right).
\ea
\eqr{eq:xiinit} then becomes
\ba
\label{eq:xid}
\xi&=&
\frac{3}{2}\alp - 3\bet + 2 \gam - \frac{5}{6}\alp\,\xi  +
      \frac{2}{3} \bet\,\xi
\nonumber\\
&+ &  {{2\,\left( -3\,{\it \alp} + 6\,{\it \bet} + {\it \alp}\,\xi
\right)}\over
{-9\,{\it \alp} + 18\,{\it \bet} + 3\,{\it \alp}\,\xi  +
           {{{\it \alp}}^2}\,\xi/2 }}
\left(\bet\, \del - 4\,{\it \alp}\,{\it \gam} +
       2\,{\it \bet}\,{\it \gam} + {\it \alp}\,{\it \gam}\,\xi   -
       {{{\it \alp}\,{{\xi }^2}}\over {12}} + \frac{\alp\xi~\tau}{12}
\right.\nonumber\\
&+ & \left.  \alp^2\left[\, -\frac{47}{8} + \frac{43\,\xi }{12} +
       \frac{2{\xi }^2}{9} -
       \frac{\xi^3}{108}~\right] +
       \alp\,\bet\left[\,\frac{29}{2}  -
       \frac{31\,\xi }{6} +
       \frac{\xi ^2}{9} ~\right] \right.\nonumber\\
&+ &\left. \bet^2\left[~- \frac{15}{2}   +
       \frac{2\,\xi}{3}\,\right] \right).
\ea
We now rearrange \eqr{eq:xid} to get an algebraic equation
for $\xi$ in terms of $\alp$, $\bet$, $\gam$ and $\del$.
In doing so, we keep only the largest
terms, keeping in mind that $\alp\ll 1$, $|\bet|\ll 1$
and $|\alp\,\xi|\ll 1$ in order that the original slow-roll expansion
is valid.
We do not assume anything about $\gam$ or $\del$, however.
Therefore, we keep the largest of all terms that include $\gam$, $\del$
and $\tau$.
(For example, we keep terms of order $\bet$, $\gam$
and $\xi$ and
neglect terms of order $\alp\bet$, $\alp\gam$,
$(\alp\xi)\gam$, $(\alp\xi)\xi$ and $(\alp\xi)^2\xi$).
The solution to \eqr{eq:xid} then is
\ba
\label{eq:xisoltn}
\xi=
\frac{1}{1-\alp\tau/18}\left(\frac{3}{2}\alp-3\bet+2\gam+\frac{2}{3}\bet\del
\right),
\ea
where we have implicitly assumed that $\tau$ does not depend on $\xi$.
If we assume that $|\alp\tau|/18\ll 1$,
$|\gam|\ll 1$ and $|\bet \del|\ll 1$,
then to first order, $\xi\simeq 3\,\alp/ 2 - 3 \bet + 2\,\gam$,
as found previously.
Note that if
$|\gam|\ga 1$, $|\bet\del|\ll |\gam|$ and $|\alp\tau|/18\ll 1$, then $\xi=
2\gam$
to first order.  In this case, $|\gam|\ll 1/(2\alp)$ in order that the
original expansion be valid.

Because we have solved for $\xi$ to first order only in general,
we can only calculate $R$ and the spectral indices to first order.
Using \eqr{eq:epeta}, the slow-roll parameters
are
$\ep=\alp/2$ and $\eta=\bet-\alp/2+\alp(\gam+\bet\del/3)/3$ to first order.
Using \eqr{eq:allfirst},
the expressions for $R$ and the spectral indices are
\ba
R&=&6.25\alp = -6.25 n_T\nonumber\\
1-n_S&=&3\alp-2\bet+\frac{2}{3}(3C-1)\alp\gam+\frac{2}{3}(C-1)\alp\bet\del
\ea
to first order.
This is similar to the first order result obtained when
$|\xi|\ll 1$, and differs only in the appearance of the
$\gam$ and $\del$ terms.
Note that when $|\bet|/\alp\gg|\gam|>1$ and $|\bet|/\alp\gg|\bet\del|$,
the first-order results are the same as when
$|\bet|/\alp\gg 1$, $|\gam|\ll 1$ and $|\bet\del|\ll 1$: $1-n_S=-2\bet$.

\end{document}